\documentclass[12pt]{article}   	
\usepackage{geometry} 
\usepackage{amsmath} 								
\usepackage{graphicx}	
\usepackage{amssymb}
\usepackage{cite}
\usepackage{color,xspace}
\usepackage{extarrows}
\usepackage{mathtools}

\usepackage[margin=1cm]{caption}

\DeclareSymbolFont{extraup}{U}{zavm}{m}{n}
\DeclareMathSymbol{\vardiamond}{\mathalpha}{extraup}{87}
\usepackage{float}
\restylefloat{table}
\allowdisplaybreaks

\def\twomat[#1,#2][#3,#4]{\left( \begin{array}{cc} #1 & #2 \\ #3 & #4 \end{array} \right)}
\def\thv[#1,#2,#3]{\left( \begin{array}{c} #1 \\ #2 \\ #3 \end{array} \right)}
\def\twv[#1,#2]{\left( \begin{array}{c} #1 \\ #2 \end{array} \right)}

\title{U(1) mixing and the Weak Gravity Conjecture }

\date{}

\begin{document}

\begin{flushright}
\end{flushright}
\begin{center}

\vspace{1cm}
{\LARGE{\bf U(1) mixing and the Weak Gravity Conjecture }}

\vspace{1cm}

\large{\bf Karim Benakli$^\spadesuit$ \let\thefootnote\relax\footnote{$^\spadesuit$kbenakli@lpthe.jussieu.fr},
Carlo Branchina$^{\vardiamond}$ \let\thefootnote\relax\footnote{$^\vardiamond$cbranchina@lpthe.jussieu.fr}
and
Ga\"etan~Lafforgue-Marmet$^\clubsuit$ \footnote{$^\clubsuit$glm@lpthe.jussieu.fr}
 \\[5mm]}

{ \sl Laboratoire de Physique Th\'eorique et Hautes Energies (LPTHE),\\ UMR 7589,
Sorbonne Universit\'e et CNRS, 4 place Jussieu, 75252 Paris Cedex 05, France.}

\end{center}
\vspace{0.7cm}

\abstract{ 
Tiny values for gauge couplings of dark photons allow to suppress their kinetic mixing with ordinary photons. We point out that the Weak Gravity Conjecture predicts consequently low ultraviolet cut-offs where new degrees of freedom might appear.  In particular, a mixing angle of $\mathcal{O}(10^{-15})$,  required in order to fit the excess reported by XENON1T, corresponds to new physics below $\mathcal{O}(100)$ TeV, thus accessible at a Future Circular Collider. We show that possible realizations are provided by compactifications with six large extra dimensions and a string scale of order $\mathcal{O}(100)$ TeV. 
}

\newpage
\setcounter{footnote}{0}

%------------------------------------------------------------------------------------------------------------							
%\section{Introduction}
%\label{introduction}

%------------------------------------------------------------------------------------------------------------

%%%%%%%%%%%%%%%%%%%%%%%%%%%%%%%%%%%%%%%%%%%%%%%%%%%%%%%%%%%%%%%%%%%%%%%%
%
%
%%%%%%%%%%%%%%%%%%%%%%%%%%%%%%%%%%%%%%%%%%%%%%%%%%%%%%%%%%%%%%%%%%%%%%%%

This short note aims to investigate some possible relations between $U(1)$ mixing and the Weak Gravity Conjecture (WGC) \cite{Vafa:2005ui,ArkaniHamed:2006dz} (for a review see e.g. \cite{Palti:2019pca}). In particular, the case of tiny mixing has witnessed a recent surge of interest  following the results  announced by XENON1T  \cite{Aprile:2020tmw}. The collaboration has reported an excess between $1$ and $7$ keV, close to the lower threshold of the experiment, with a peak around $2 - 3$ keV.  The significance of this excess could melt with a re-analysis of the signal. This could either follow from an accumulation of more data or from more thorough searches for evidences of contamination of the apparatus by some impurities as for the tritium hypothesis suggested by the collaboration in \cite{Aprile:2020tmw}. In the meanwhile, the possibility that it could be a signal of new physics does not seem excluded.

A possible fit of the data in terms of dark photons coupled to the Standard Model (SM) through a kinetic mixing portal \cite{Holdom:1985ag,Okun:1982xi,delAguila:1988jz,Dienes:1996zr} was analyzed in \cite{An:2020bxd,Alonso-Alvarez:2020cdv,Choi:2020udy}.
While solar emitted dark photons are not favored, a scenario where light dark photons with masses of $2 - 4$ keV are absorbed by the xenon seems to correctly reproduce the excess, though with a reduced significance due to a look elsewhere effect. This can be achieved for a tiny visible-dark photon kinetic mixing parameter in the range

\begin{equation}
\label{epsilonconstraint}
\epsilon\simeq \mathcal{O}(10^{-16}-10^{-15})    
\end{equation}
which is in agreement with the upper bound limit given by XENON1T on $\epsilon$, that also claims a $3\sigma$ significance for a $2.3$ keV dark photon over the background. This was argued in \cite{Nakayama:2020ikz} to lead to the correct result for the dark photon relic density. The dark photon in XENON1T can also appear as a vector portal for fermionic or bosonic dark matter, where, depending on the model, the mixing can take different values. In \cite{Baek:2020owl,Ko:2020gdg,Zu:2020idx}, mixing parameters of order $\epsilon \simeq \mathcal{O}(10^{-4}), \mathcal{O}(10^{-7})$ or $\mathcal{O}(10^{-10})$, with, respectively, an order  $\mathcal{O}(\mathrm{GeV})$ massive dark photon in the first two cases and a massless one in the last one have been advocated. 
We discuss below a possible origin of such mixing parameter, especially for challenging tiny values, where we find that the WGC allows to hope for an accompanying  signal at collider experiments.

We focus on the sector of the low energy effective field theory describing the $U(1)$ gauge groups representations and interactions. One of the two, $U(1)_v$, is called visible as we have in mind hypercharge or electromagnetism. Another, $U(1)_d$, corresponds  to an extra factor we call "dark" $U(1)$, having in mind an hidden sector. It is straightforward to generalize to cases with more abelian gauge groups. The associated gauge fields and gauge fields strengths are denoted as $A^{\mu}_{(v)}, F^{\mu \nu}_{(v)}$ and $A^{\mu}_{(d)}, F^{\mu \nu}_{(d)}$, respectively. The corresponding two-derivative Lagrangian reads:
\begin{equation}
\mathcal{L} \supset -\frac{1}{4} F^{\mu \nu}_{(v)} F_{(v) \mu \nu} -\frac{1}{4} F^{\mu \nu}_{(d)} F_{(d) \mu \nu} -\frac{\epsilon_{vd}}{2} F^{\mu \nu}_{(v)} F_{(d) \mu \nu}
+ g_v J^{\mu}_{(v)} A_{(v) \mu} + g_d J^{\mu}_{(d)} A_{(d) \mu} \, .
\label{Lagrangian}
\end{equation}

For massless  visible and dark photons, this mixing in the two-derivative Lagrangian can be eliminated by performing the appropriate rotation. When the $U(1)_d$ gauge boson acquires a mass, through a Stueckelberg or Higgs mechanisms, the mixing has physical implications. The visible and dark photons couple in the new basis to the currents $J^{\mu}_v$ and $J^{\mu}_d$ through:
\begin{equation}
\mathcal{L} \supset \left[ \frac{g_d}{\sqrt{1- \epsilon_{vd}^2}}J^{\mu}_{(d)} - \frac{  \epsilon_{vd} g_v}{\sqrt{1- \epsilon_{vd}^2}}J^{\mu}_{(v)} \right] A_{(d) \mu} + g_v J^{\mu}_{(v)} A_{(v) \mu} \, ,
\end{equation}
thus implying that the visible matter is charged under the dark gauge symmetry with charge $\sim \epsilon_{vd} g_v$.

It is most natural to assume that the dark $U(1)$ mass and mixing  vanish in the fundamental theory at the ultra-violet (UV) cut-off and are generated at lower energies. The mixing can be generated at one  loop by states with masses $m_i$ and charges $(q_v^{(i)}, q_d^{(i)})$ under $(U(1)_v , U(1)_d) $. It is then given by:
\begin{equation}
\epsilon_{v d} =\frac{g_v g_d  }{16 \pi^2} \sum_i q_v^{(i)} q_d^{(i)} \ln {\frac{ m_i^2}{{\mu^2}}}, 
\label{epsilonLog}
\end{equation}
where $\mu^2$ is the renormalisation scale. In the case of the hyper-charge $U(1)_v \equiv U(1)_Y $ we have $g_v=g'$ and $q_v^{(i)}=Y^{(i)}$, while $g_v=g' \cos{\theta_w}$ and $q_v^{(i)}=q_{em}^{(i)}$  the electrical charge for $U(1)_v \equiv U(1)_{em}$.

In order to generate such a small mixing as the one required by XENON1T, we either require the dark photon coupling to be appropriately small, a cancellation in the one-loop logarithms,  or appeal to higher order non-renormalisable operators. The cancellation can be partial, for instance between particles with (order one) charges $(q_v^{(i)}, q_d^{(i)})$ and $(q_v^{(j)}, q_d^{(j)}= - q_d^{(i)})$ and masses $m_i$ and $m_j$ with $m_j=m_i + \Delta m_{ij}$. For $\Delta m_{ij} \ll m_i$, we have an approximation:
\begin{equation}
\epsilon_{v d}\sim \frac{g_v g_d  }{16 \pi^2}  {\frac{ \Delta m_{ij}}{{m_i}}} \, .
\end{equation}
For complete cancellation, this one loop contribution is replaced by higher loop ones. However, gravitational loops are expected to show up at some order and lead to a lower bound. It was shown in \cite{Gherghetta:2019coi} that this is expected at six loop order giving rise to an 
$\epsilon_{v d} \gtrsim \mathcal{O}(10^{-13})$ for a {\it bona fide} four dimensional theory. We shall discuss below the first alternative of a tiny dark sector coupling.

We start by considering an abelian gauge symmetry $U(1)$ with gauge coupling $g$. The Weak Gravity Conjecture requires the presence of at least one state with a mass:
\begin{equation}
m^2  \le 2 {g^2 q^2} \, \,  {M_P^2} 
\label{mqratio}
\end{equation}
where we use use natural units $\hbar = c =1$ and $M_P ={\sqrt{8 \pi G}} \sim 2.4  \times 10^{18}$ GeV is the reduced Planck mass. Obviously this is satisfied by all the Standard Model states for the hypercharge/electromagnetic gauge symmetry. 
This bound was generalised to the case of multiple $U(1)$s by the replacement $q^2 \rightarrow \sum_i q_i^2$ in \eqref{mqratio} \cite{Cheung:2014vva}.  In theories with supersymmetry, extra-dimensions can furnish a set of BPS states (Kaluza-Klein modes or solitonic objects as branes) that saturate this bound. Their masses will receive corrections after supersymmetry breaking but some should still satisfy the bound. 

An explicit model is required for an exact computation of the mixing given by (\ref{epsilonLog}). Here, we would like to comment on possible states that will contribute to the sum. Following  the Completeness Hypothesis of \cite{Polchinski:2003bq},  all the sets of $U(1)$ charges are present in the theory. It was suggested in \cite{Heidenreich:2015nta} that all the spots of the charge lattice are occupied by  super-extremal states, i.e. satisfying the WGC. This Lattice Weak Gravity Conjecture was shown to be too strong; it does not to hold in Kaluza-Klein theories within generic compactifications. It was subsequently replaced by the sub-Lattice Weak Gravity Conjecture \cite{Heidenreich:2016aqi}. This states instead that a super-extremal particle should exist only at every spot in a finite index sub-lattice of the full charge lattice. Later, causality and analyticity constraints for lower dimensional theories, obtained through dimensional reduction, has been used in \cite{Andriolo:2018lvp} to propose the Tower Weak Gravity Conjecture: there is an infinite tower of states satisfying the bounds of the WGC. These states will contribute to generating a mixing between the $U(1)$s. For two $U(1)$s, the masses of the charged particles can be expressed as $ m=c\sqrt{(g_vq_v)^2+(g_dq_d)^2}$, where $c<1$ is a state-dependent constant. For the following discussion, we use two integers $i$ and $j$ for the visible and dark charge of the particle respectively, as would be for quantized charges forming a lattice in charge space. Eq. \eqref{epsilonLog} then becomes in this scheme
\begin{equation}
\epsilon_{v d} =\frac{g_v g_d  }{16 \pi^2} \sum_{i,j} q_i q_j \ln \left(\frac{ c^{(i,j)}\left[(g_v q_i)^2+(g_d q_j)^2\right]}{\mu^2}\right)
\end{equation}
Though the number of states is infinite, we include in the loop only states below the cut-off. If a particle with  charge $(q_i,q_j)$ is in the spectrum, there are also particles with charge $(q_i,-q_j)$, $(-q_i,q_j)$ and $(-q_i,-q_j)$ giving
 \begin{equation}
 \epsilon_{v d} \simeq \frac{g_v g_d  }{16 \pi^2} \sum_{i,j} q_i q_j  \ln \left(\frac{ c^{(i,j)}c^{(-i,-j)}}{c^{(-i,j)}c^{(i,-j)}}\right)
\end{equation}  
which as a result of the diverse cancellation between different contributions, could remain small (for typical sizes, see for example discussion in \cite{Dienes:1996zr}). 

The most relevant facet of the WGC for this work is the prediction of an  ultraviolet cut-off scale for the effective field theory at $\Lambda_{UV} \lesssim g { M_{Pl}}$. This was dubbed as the magnetic weak gravity conjecture in \cite{ArkaniHamed:2006dz} and, in the weak coupling limit $g\to 0$, it predicts the absence of global symmetries in quantum gravity \cite{Polchinski:2003bq,Banks:2010zn,Harlow:2018tng}. For electromagnetic or hypercharge gauge coupling, the cut-off scale set by the WGC remains close to the Planck scale.

We generalize here, as done by \cite{Heidenreich:2017sim}, this requirement to the case with multiple $U(1)$ gauge groups by requiring that none of the gauge symmetry factors should turn into a continuous global symmetry by taking the corresponding coupling to vanish. This implies that a tiny value of the dark photon gauge coupling, introduced to make the mixing tiny, require  a UV cut-off at most of order $\Lambda_{UV} \lesssim g_dM_P$. This is sensibly lower than $M_P$ and could have important consequences in phenomenology and cosmology.

Starting from (\ref{epsilonLog}), we identify the visible photon with the SM photon, i.e. $g_v = e \sim 0.3$, and the logarithm to be $\mathcal{O}(1 - 10)$, then
\begin{equation}
\epsilon_{v d}\sim \frac{g_v g_d  }{16 \pi^2}  \sim \mathcal{O}(10^{-3} - 10^{-2}) g_d \qquad  \Rightarrow \qquad   g_d \sim \mathcal{O}(10^{2} - 10^{3}) \epsilon_{v d} 
\label{g_d from epsilon}
\end{equation}
Per se, the WGC does not provide information on the new physics required at $\Lambda_{UV} \lesssim g_d { M_{Pl}}$. A simple possibility is that the $U(1)_d$ becomes part of some non-abelian gauge group $SU(2)_D$ with field strength $F^{\mu \nu}_{(D)} $ broken by a vacuum expectation value (v.e.v) of  $\left\langle\Sigma\right\rangle=v \simeq  \Lambda_{UV}/g_d$ of a field in the adjoint representation. One could then induce a contribution  $\epsilon_{v d}^{NR}$ to the kinetic mixing through the effective non-renormalizable operator (see e.g. \cite{Essig:2009nc}):
\begin{equation}
\frac{c^{NR}}{M_P} Tr \left[ \Sigma F^{\mu \nu}_{(D)}  \right]  F_{(v) \mu \nu} \quad \Rightarrow  \quad \epsilon_{v d}^{NR} \simeq \frac{c^{NR} \, \,  v}{ M_P}  
\end{equation}
where $c^{NR}$ is a constant. For this contribution to remain sub-leading, we require:
\begin{equation}
c^{NR} \, \,  v  \lesssim \epsilon_{v d} M_P  \quad  \Rightarrow   \quad  c^{NR} \, \,  v   \lesssim 10^{-3} \, \,  g_d \, \,  { M_P} , 
\end{equation}
which  for $\epsilon_{v d} \sim 10^{-15}$ gives $ c^{NR} \, \,  v   \lesssim $ TeV.

Kinetic mixing might also arise from D-terms in supersymmetric theories through effective operators  \cite{Benakli:2009mk,Goodsell:2009xc}:
\begin{equation}
\frac{D^2}{\Lambda_D^4}  F^{\mu \nu}_{(d)}  F_{(v) \mu \nu}  
\end{equation}
that are expected to be very small. For example, they can be suppressed by the value of the ratio SM Higgs v.e.v over the scale $\Lambda_D$ for hypercharge $D$-term or through powers of the dark sector coupling for the dark $U(1)$ D-term.

In the following we will use \eqref{g_d from epsilon} to compute $g_d$ from $\epsilon$. A value of $\epsilon_{v d} \sim 10^{-15}$ as in \eqref{epsilonconstraint} would require $g_d  \sim  \mathcal{O}(10^{-13} - 10^{-12})$. The WGC implies then that the theory has a UV cut-off:
\begin{equation}
\Lambda_{UV} \lesssim g_d { M_{P}} \sim  \mathcal{O}(10^{2} - 10^{3}) {\rm TeV}.
\label{LambdaUV}
\end{equation}
Therefore, new physics must appear below energies of order $\mathcal{O}(100)$ TeV. Such physics could be accessible at future experiments at collider, such as the 100 TeV Future Circular Collider (FCC).

Following the SLP, such a scenario is consistent with quantum gravity only if it could arise from a string theory model. We will discuss now one possible venue for realizing this UV completion in a string theory. We do not attempt an explicit string model building which is beyond the scope of this work.  We contemplate the possibility that a hierarchy $g_d \ll g_v$ is obtained through the suppression of $g_d$ by the volume of the internal compactified space. More precisely, we consider a scenario where we start from ten-dimensional type IIB string theory compactified on a six-dimensional space of volume $V_6 \equiv (2 \pi R)^6$. The four-dimensional reduced Planck mass $M_P$ is related to the string scale mass $M_s$ and string coupling $g_s$ through (multiplied by $2$ for type I strings):
\begin{equation}
M_P^2 = \frac{R^6 M_s^6} {2 \pi g_s^2} M_s^2
\end{equation}
The visible $U(1)_v$ is taken to live on a D5-brane wrapping a small two-dimensional cycle of approximate string size of volume $ (2 \pi r)^2 \gtrsim 4 \pi^2 M_s^{-2}$. Then, the visible coupling reads:
\begin{equation}
g_v^2= \frac {2 \pi g_s} {r^2 M_s^2} \simeq 2 \pi g_s
\end{equation}
The dark $U(1)_d$ is instead chosen to live on a D9-brane wrapping the whole six-dimensional compact space and its gauge coupling is given by:
\begin{equation}
g_d^2= \frac {2 \pi g_s} {R^6 M_s^6} 
\end{equation}
Then, we get:
\begin{equation}
\epsilon_{v d} \sim \frac{g_v g_d  }{16 \pi^2}  \sim  \frac { g_s} {8 \pi R^3 M_s^3}   \quad \Rightarrow \quad  \epsilon_{v d} \sim \frac{1  }{ \sqrt{128 \pi^3}} \frac{M_s }{ M_P}  \sim 10^{-2} \frac{M_s }{ M_P} 
\end{equation}
thus
\begin{equation}
\epsilon_{v d} \sim 10^{-15}   \quad \Rightarrow \quad  {M_s } \sim  \mathcal{O}(100) {\rm TeV}
\end{equation}

This is merely two orders of magnitude above the proposals of TeV strings for solving the hierarchy problem \cite{Antoniadis:1990ew,Antoniadis:1993jp,Antoniadis:1994yi,ArkaniHamed:1998rs,Antoniadis:1998ig,Lykken:1996fj,Dienes:1998vh,Dienes:1998vg,Benakli:1998pw,Randall:1999ee}. Note that our analysis is similar to the analysis performed in \cite{Goodsell:2009xc}. However there is a notable difference in that we impose that the  $U(1)_d$ dark propagates in the whole large dimensions, thus six in this example, therefore we have considered D5-D9 branes instead of D3-D7 , leading to different results, and in particular allowing smaller values of the mixing. The Dp-D(p-4) set-up is enforced by supersymmetry, but, in the case of low string scale, we could have taken instead, without change in our results, a non-supersymmetric configuration of D3-D9 branes, our world being non-supersymmetric at least up to TeV energy scales. However, one should keep in mind that some of the non-supersymmetric configurations tend to fall in the Swampland \cite{Ooguri:2006in}.

The above scenario implies the appearance of large extra-dimensions at a scale of order:
\begin{equation}
\frac{1}{R} = \left( \frac{M_s}{\sqrt{8\pi}M_P} \frac{1}{\alpha_{YM}} \right)^{1/3} M_s
\end{equation}
where we have identified the tree-value of the SM gauge couplings as $\alpha_{YM} = g_s / 2$. Taking an approximate value for $\alpha_{YM} \sim 1/ 25$, we get 
\begin{equation}
\frac{1}{R}  \sim  \mathcal{O}(10) \,{\rm GeV}
\end{equation}
Though these values of the compactification energy scale might seem low, they are not experimentally excluded. Gauge bosons propagate in these extra dimensions, in addition to the gravitons. However, in contrast to the case in \cite{Antoniadis:1990ew,Antoniadis:1993jp,Antoniadis:1994yi,Antoniadis:2000vd,Accomando:1999sj,Antoniadis:1999bq}, these are Kaluza-Klein excitations of the dark $U(1)$ with tiny couplings. It is the production of a huge number of them that will compensate the coupling strong suppression. They can be observed as missing energy at collider experiments in particular at a 100 TeV collider.

We can express the string mass scale and the compactification radius as a fonction of $g_d$ and $M_P$ :
\begin{equation}
    M_s \sim \sqrt{g_s}g_dM_P \quad \mathrm{and} \quad \frac{1}{R} \sim \frac{g_d^{\frac{4}{3}}}{(8\pi)^\frac{1}{6}}M_P \lesssim g_dM_P
\end{equation}

The most stringent bound on $\epsilon$ today is $\epsilon \sim \mathcal O(10^{-16})$ (see e.g. \cite{Fabbrichesi:2020wbt}), with a mass for the dark photon around the keV. For this case, one obtains for the string mass scale $M_s \sim 10^4$ GeV, and $\frac{1}{R} \sim 0.1-1$ GeV. Smaller values of $\epsilon$ cannot be obtained through our simple large extra-dimension setup as they will conflict with the current experimental limit on the string scale. For different values of the dark photon mass, the constraints on $\epsilon$ are weaker, and consequently the extra-dimension and string mass scales are set at higher energies. Taking the three values mentioned above we can have for $\epsilon \sim10^{-10}$, $\epsilon \sim 10^{-7} $ and $\epsilon \sim 10^{-4}$, respectively, a string mass scale and an inverse compactification radius of order $M_s \sim 10^{10}$ GeV  and $\frac{1}{R} \sim 10^7$ GeV, $M_s \sim 10^{13} $ GeV and $\frac{1}{R} \sim 10^{11}$ GeV, and finally $M_s \sim 10^{16} $ GeV and $\frac{1}{R} \sim 10^{15}$  GeV. 
The intermediate scale $\sim 10^{11} $ GeV has diverse motivations \cite{Burgess:1998px,Benakli:1998pw}. It also corresponds to the energy where the SM quartic Higgs coupling vanish, thus a scale where new degrees of freedom might be expected. Though we cannot proceed to the same string embedding as we have done above, for kinetic mixing as small as $\epsilon\sim 10^{-23}$ the WGC requires new physics around the scale $\Lambda\sim 1-10$ MeV, that could then in turn be constrained by the Big Bang Nucleosynthesis.
 
In our way to generate the large hierarchy between the two couplings $g_v$ and $g_d$, we have assumed the existence of small cycle with a size of order of the string scale inside six large compact dimensions. This is not the case in the simplest toroidal compactifications and requires some warping. Thus, the KK excitations of the dark photon are not expected in general to exhibit the same spectrum as in the simplest case. However, assuming that the rough behaviour of the density of states goes with the energy as $E^6 / M_s^6$, a sizable value of the effective coupling between SM states and the dark photons is reached only at energies of the order of $M_s$.   

In most phenomenological applications, the dark $U(1)$ is massive. The WGC in \cite{ArkaniHamed:2006dz} concerns massless $U(1)$ gauge bosons. For instance, in the case of a massive $U(1)$, the charge is not conserved and there is no problem of remnants as charged black holes decay. However, one may argue that if the weak gravity states masses $m_{WGC}$ are much bigger than the dark photon mass $m_{\gamma_d}$, the massive case is a (Higgs) phase of the same theory and remains in the landscape. Moreover, the comparison of gravity and gauge forces should be done at energies of order $m_{WGC}$ and makes sense in the region $m_{\gamma_d}\ll m_{WGC}$. Finally, we have explicitly illustrated the WGC prediction for the UV cut-off of the theory by a type IIB string scenario that we do not expect to break down because of an infrared Higgsing of the $U(1)$.  In fact,  \cite{Heidenreich:2017sim} have argued, through the explicit investigations of the properties of the WGC charge lattice, that the bounds used here on the mass, combination of charges ratios and  ultraviolet cut-off of the theory remain true. A detailed discussion of the expected masses for the dark photon in different string settings is provided elsewhere \cite{Anchordoqui:2020tlp}.

To conclude, we would like to stress that the main aim of this work is not to add to the plethora of XENON1T analysis and interpretation, but to point out the amusing coincidence that the observation of kinetic mixing between ordinary and dark photon would suggest new physics at scales that should be probed by a future collider.

%%%%%%%%%%%%%%%%%%%%%%%%%%%%%%%%%%%%%%%%%%%%%%%%%

%\vskip.1in
\noindent
\section*{Acknowledgments}
K.B. thanks Ignatios Antoniadis and Marco Cirelli for useful discussions. We acknowledge the support of  the Agence Nationale de Recherche under grant ANR-15-CE31-0002 ``HiggsAutomator''.

%\newpage

\noindent

\providecommand{\href}[2]{#2}\begingroup\raggedright\endgroup


\begin{thebibliography}{10}
	
	%\cite{Vafa:2005ui}
	\bibitem{Vafa:2005ui}
	C.~Vafa,
	\emph{The String landscape and the swampland},
	[hep-th/0509212].
	%%CITATION = HEP-TH/0509212;%%
	
	 %\cite{ArkaniHamed:2006dz}
	\bibitem{ArkaniHamed:2006dz}
	N.~Arkani-Hamed, L.~Motl, A.~Nicolis and C.~Vafa,
	\emph{The String landscape, black holes and gravity as the weakest force},
	JHEP {\bf 0706} (2007) 060
	% doi:10.1088/1126-6708/2007/06/060
	[hep-th/0601001].
	%%CITATION = doi:10.1088/1126-6708/2007/06/060;%%
	
	%\cite{Palti:2019pca}
	\bibitem{Palti:2019pca}
	E.~Palti,
    \emph{The Swampland: Introduction and Review},
	Fortsch.\ Phys.\  {\bf 67} (2019) no.6,  1900037
    %doi:10.1002/prop.201900037
	[arXiv:1903.06239 [hep-th]].
	%%CITATION = doi:10.1002/prop.201900037;%%


    %\cite{Aprile:2020tmw}
    \bibitem{Aprile:2020tmw}
    E.~Aprile \textit{et al.} [XENON],
    \emph{Observation of Excess Electronic Recoil Events in XENON1T},
    [arXiv:2006.09721 [hep-ex]].

    
    %\cite{Holdom:1985ag}
    \bibitem{Holdom:1985ag}
    B.~Holdom,
    \emph{Two U(1)'s and Epsilon Charge Shifts},
    Phys. Lett. B \textbf{166} (1986), 196-198
    %doi:10.1016/0370-2693(86)91377-8
    

    %\cite{Okun:1982xi}
    \bibitem{Okun:1982xi}
    L.~B.~Okun,
    \emph{Limits of electrodynamics: paraphotons?},
    Sov. Phys. JETP \textbf{56} (1982), 502
    ITEP-48-1982.


    %\cite{delAguila:1988jz}
    \bibitem{delAguila:1988jz}
    F.~del Aguila, G.~D.~Coughlan and M.~Quiros,
    \emph{Gauge Coupling Renormalization With Several U(1) Factors},
    Nucl. Phys. B \textbf{307} (1988), 633
    %doi:10.1016/0550-3213(88)90266-0


    %\cite{Dienes:1996zr}
    \bibitem{Dienes:1996zr}
    K.~R.~Dienes, C.~F.~Kolda and J.~March-Russell,
    \emph{Kinetic mixing and the supersymmetric gauge hierarchy},
    Nucl. Phys. B \textbf{492} (1997), 104-118
    %doi:10.1016/S0550-3213(97)00173-9
    [arXiv:hep-ph/9610479 [hep-ph]].



%%%%%%%%%%%%%%%%%%%%%%%%%%%%%%%%%  Xenon and Dark Photon 
	
    %\cite{An:2020bxd}
    \bibitem{An:2020bxd}
    H.~An, M.~Pospelov, J.~Pradler and A.~Ritz,
    \emph{New limits on dark photons from solar emission and keV scale dark matter},
    [arXiv:2006.13929 [hep-ph]].
    
    %\cite{Alonso-Alvarez:2020cdv}
	\bibitem{Alonso-Alvarez:2020cdv}
	G.~Alonso-Álvarez, F.~Ertas, J.~Jaeckel, F.~Kahlhoefer and 		L.~J.~Thormaehlen,
	\emph{Hidden Photon Dark Matter in the Light of XENON1T and Stellar Cooling},
	[arXiv:2006.11243 [hep-ph]].
	
	
	%\cite{Choi:2020udy}
	\bibitem{Choi:2020udy}
	G.~Choi, M.~Suzuki and T.~T.~Yanagida,
	\emph{XENON1T Anomaly and its Implication for Decaying Warm Dark Matter},
	[arXiv:2006.12348 [hep-ph]].
	
	
	%\cite{Nakayama:2020ikz}
	\bibitem{Nakayama:2020ikz}
	K.~Nakayama and Y.~Tang,
	\emph{Gravitational Production of Hidden Photon Dark Matter in light of the XENON1T Excess},
	[arXiv:2006.13159 [hep-ph]]. 
    
   %\cite{Baek:2020owl}
	\bibitem{Baek:2020owl}
	S.~Baek, J.~Kim and P.~Ko,
	\emph{XENON1T excess in local $Z_2$ DM models with light dark 			sector},
	[arXiv:2006.16876 [hep-ph]].
	 
     
	%\cite{Ko:2020gdg}
	\bibitem{Ko:2020gdg}
	P.~Ko and Y.~Tang,
	\emph{Semi-annihilating $Z_3$ Dark Matter for XENON1T Excess},
	[arXiv:2006.15822 [hep-ph]].
	
	    %\cite{Zu:2020idx}
	\bibitem{Zu:2020idx}
	L.~Zu, G.~W.~Yuan, L.~Feng and Y.~Z.~Fan,
	\emph{Mirror Dark Matter and Electronic Recoil Events in 		XENON1T},
	[arXiv:2006.14577 [hep-ph]].
	    


	
	%\cite{Gherghetta:2019coi}
    \bibitem{Gherghetta:2019coi}
    T.~Gherghetta, J.~Kersten, K.~Olive and M.~Pospelov,
    \emph{Evaluating the price of tiny kinetic mixing},
    Phys. Rev. D \textbf{100} (2019) no.9, 095001
    %doi:10.1103/PhysRevD.100.095001
    [arXiv:1909.00696 [hep-ph]].
    
     %\cite{Cheung:2014vva}
    \bibitem{Cheung:2014vva}
    C.~Cheung and G.~N.~Remmen,
    \emph{Naturalness and the Weak Gravity Conjecture},
    Phys. Rev. Lett. \textbf{113} (2014), 051601
    %doi:10.1103/PhysRevLett.113.051601
    [arXiv:1402.2287 [hep-ph]].
	
%	%\cite{Kim:2019ths}
%    \bibitem{Kim:2019ths}
%    H.~C.~Kim, H.~C.~Tarazi and C.~Vafa,
%    \emph{Four Dimensional $\mathbf{\mathcal{N}=4}$ SYM and the Swampland},
%    [arXiv:1912.06144 [hep-th]].


	
%%%%%%%%%%%%%%%%% absence of global symmetries
    %\cite{Polchinski:2003bq}
    \bibitem{Polchinski:2003bq}
    J.~Polchinski,
    \emph{Monopoles, duality, and string theory},
    Int. J. Mod. Phys. A \textbf{19S1} (2004), 145-156
    %doi:10.1142/S0217751X0401866X
    [arXiv:hep-th/0304042 [hep-th]].
    %78 citations counted in INSPIRE as of 06 Jul 2020

	%\cite{Heidenreich:2015nta}
	\bibitem{Heidenreich:2015nta}
	B.~Heidenreich, M.~Reece and T.~Rudelius,
	\emph{Sharpening the Weak Gravity Conjecture with Dimensional 			Reduction},
	JHEP \textbf{02} (2016), 140
	%doi:10.1007/JHEP02(2016)140
	[arXiv:1509.06374 [hep-th]].
	%130 citations counted in INSPIRE as of 04 Nov 2020
	
	%\cite{Heidenreich:2016aqi}
	\bibitem{Heidenreich:2016aqi}
	B.~Heidenreich, M.~Reece and T.~Rudelius,
	\emph{Evidence for a sublattice weak gravity conjecture},
	JHEP \textbf{08} (2017), 025
	%doi:10.1007/JHEP08(2017)025
	[arXiv:1606.08437 [hep-th]].
	%110 citations counted in INSPIRE as of 04 Nov 2020

	%\cite{Andriolo:2018lvp}
	\bibitem{Andriolo:2018lvp}
	S.~Andriolo, D.~Junghans, T.~Noumi and G.~Shiu,
	\emph{A Tower Weak Gravity Conjecture from Infrared 				Consistency},
	Fortsch. Phys. \textbf{66} (2018) no.5, 1800020
	%doi:10.1002/prop.201800020
	[arXiv:1802.04287 [hep-th]].
	%67 citations counted in INSPIRE as of 04 Nov 2020


    %\cite{Banks:2010zn}
    \bibitem{Banks:2010zn}
    T.~Banks and N.~Seiberg,
    \emph{Symmetries and Strings in Field Theory and Gravity},
    Phys. Rev. D \textbf{83} (2011), 084019
    %doi:10.1103/PhysRevD.83.084019
    [arXiv:1011.5120 [hep-th]].
    %404 citations counted in INSPIRE as of 06 Jul 2020
    
    %\cite{Harlow:2018tng}
    \bibitem{Harlow:2018tng}
    D.~Harlow and H.~Ooguri,
    \emph{Symmetries in quantum field theory and quantum gravity},
    [arXiv:1810.05338 [hep-th]].
    %86 citations counted in INSPIRE as of 06 Jul 2020



%%%%%%%%%%%%%%%%%%%%%%%%%%%%%%%%%%%%%%%%%%%%%%%%%%%

    %\cite{Heidenreich:2017sim}
    \bibitem{Heidenreich:2017sim}
    B.~Heidenreich, M.~Reece and T.~Rudelius,
    \emph{The Weak Gravity Conjecture and Emergence from an Ultraviolet Cutoff},
    Eur. Phys. J. C \textbf{78} (2018) no.4, 337
    %doi:10.1140/epjc/s10052-018-5811-3
    [arXiv:1712.01868 [hep-th]].
    %51 citations counted in INSPIRE as of 05 Jul 2020
    
    %\cite{Essig:2009nc}
    \bibitem{Essig:2009nc}
    R.~Essig, P.~Schuster and N.~Toro,
    \emph{Probing Dark Forces and Light Hidden Sectors at Low-Energy e+e- Colliders},
    Phys. Rev. D \textbf{80} (2009), 015003
    %doi:10.1103/PhysRevD.80.015003
    [arXiv:0903.3941 [hep-ph]].
    
     %\cite{Benakli:2009mk}
    \bibitem{Benakli:2009mk}
    K.~Benakli and M.~D.~Goodsell,
    \emph{Dirac Gauginos and Kinetic Mixing},
    Nucl. Phys. B \textbf{830} (2010), 315-329
    %doi:10.1016/j.nuclphysb.2010.01.003
    [arXiv:0909.0017 [hep-ph]].

   %\cite{Goodsell:2009xc}
    \bibitem{Goodsell:2009xc}
    M.~Goodsell, J.~Jaeckel, J.~Redondo and A.~Ringwald,
    \emph{Naturally Light Hidden Photons in LARGE Volume String Compactifications},
    JHEP \textbf{11} (2009), 027
    %doi:10.1088/1126-6708/2009/11/027
    [arXiv:0909.0515 [hep-ph]].

     %%%%%%%%%%%%%%%%%%%%%%%%%%%%%%%%%%%%%%%%%%%



    %\cite{Antoniadis:1990ew}
    \bibitem{Antoniadis:1990ew}
    I.~Antoniadis,
    \emph{A Possible new dimension at a few TeV},
    Phys. Lett. B \textbf{246} (1990), 377-384
    %doi:10.1016/0370-2693(90)90617-F
    

    %\cite{Antoniadis:1993jp}
    \bibitem{Antoniadis:1993jp}
    I.~Antoniadis and K.~Benakli,
    \emph{Limits on extra dimensions in orbifold compactifications of superstrings},
    Phys. Lett. B \textbf{326} (1994), 69-78
    %doi:10.1016/0370-2693(94)91194-0
    [arXiv:hep-th/9310151 [hep-th]].


    %\cite{Antoniadis:1994yi}
    \bibitem{Antoniadis:1994yi}
    I.~Antoniadis, K.~Benakli and M.~Quiros,
    \emph{Production of Kaluza-Klein states at future colliders},
    Phys. Lett. B \textbf{331} (1994), 313-320
    %doi:10.1016/0370-2693(94)91058-8
    [arXiv:hep-ph/9403290 [hep-ph]].

    %\cite{ArkaniHamed:1998rs}
    \bibitem{ArkaniHamed:1998rs}
    N.~Arkani-Hamed, S.~Dimopoulos and G.~R.~Dvali,
    \emph{The Hierarchy problem and new dimensions at a millimeter},
    Phys. Lett. B \textbf{429} (1998), 263-272
    %doi:10.1016/S0370-2693(98)00466-3
    [arXiv:hep-ph/9803315 [hep-ph]].


    %\cite{Antoniadis:1998ig}
    \bibitem{Antoniadis:1998ig}
    I.~Antoniadis, N.~Arkani-Hamed, S.~Dimopoulos and G.~R.~Dvali,
    \emph{New dimensions at a millimeter to a Fermi and superstrings at a TeV},
    Phys. Lett. B \textbf{436} (1998), 257-263
    %doi:10.1016/S0370-2693(98)00860-0
    [arXiv:hep-ph/9804398 [hep-ph]].

   %\cite{Lykken:1996fj}
    \bibitem{Lykken:1996fj}
    J.~D.~Lykken,
    \emph{Weak scale superstrings},
    Phys. Rev. D \textbf{54} (1996), 3693-3697
    %doi:10.1103/PhysRevD.54.R3693
    [arXiv:hep-th/9603133 [hep-th]].
    
 
    %\cite{Dienes:1998vh}
    \bibitem{Dienes:1998vh}
    K.~R.~Dienes, E.~Dudas and T.~Gherghetta,
    \emph{Extra space-time dimensions and unification},
    Phys. Lett. B \textbf{436} (1998), 55-65
    %doi:10.1016/S0370-2693(98)00977-0
    [arXiv:hep-ph/9803466 [hep-ph]].



    %\cite{Dienes:1998vg}
    \bibitem{Dienes:1998vg} 
    K.~R.~Dienes, E.~Dudas and T.~Gherghetta,
    \emph{Grand unification at intermediate mass scales through extra dimensions},
    Nucl. Phys. B \textbf{537} (1999), 47-108
    %doi:10.1016/S0550-3213(98)00669-5
    [arXiv:hep-ph/9806292 [hep-ph]].
    
    %\cite{Benakli:1998pw}
    \bibitem{Benakli:1998pw}
    K.~Benakli,
    \emph{Phenomenology of low quantum gravity scale models},
    Phys. Rev. D \textbf{60} (1999), 104002
    %doi:10.1103/PhysRevD.60.104002
    [arXiv:hep-ph/9809582 [hep-ph]].

    %\cite{Randall:1999ee}
    \bibitem{Randall:1999ee}
    L.~Randall and R.~Sundrum,
    \emph{A Large mass hierarchy from a small extra dimension},
    Phys. Rev. Lett. \textbf{83} (1999), 3370-3373
    %doi:10.1103/PhysRevLett.83.3370
    [arXiv:hep-ph/9905221 [hep-ph]].
    
   
  
	%\cite{Ooguri:2006in}
	\bibitem{Ooguri:2006in}
	H.~Ooguri and C.~Vafa,
	\emph{On the Geometry of the String Landscape and the Swampland},
	Nucl.\ Phys.\ B {\bf 766} (2007) 21
    %doi:10.1016/j.nuclphysb.2006.10.033
	[hep-th/0605264].
	%%CITATION = doi:10.1016/j.nuclphysb.2006.10.033;%%
	
	 %\cite{Antoniadis:2000vd}
    \bibitem{Antoniadis:2000vd}
    I.~Antoniadis and K.~Benakli,
    \emph{Large dimensions and string physics in future colliders},
    Int. J. Mod. Phys. A \textbf{15} (2000), 4237-4286
    %doi:10.1016/S0217-751X(00)00217-0
    [arXiv:hep-ph/0007226 [hep-ph]].
    %97 citations counted in INSPIRE as of 06 Jul 2020

    %\cite{Accomando:1999sj}
    \bibitem{Accomando:1999sj}
    E.~Accomando, I.~Antoniadis and K.~Benakli,
    \emph{Looking for TeV scale strings and extra dimensions},
    Nucl. Phys. B \textbf{579} (2000), 3-16
    %doi:10.1016/S0550-3213(00)00123-1
    [arXiv:hep-ph/9912287 [hep-ph]].
    %157 citations counted in INSPIRE as of 06 Jul 2020

    %\cite{Antoniadis:1999bq}
    \bibitem{Antoniadis:1999bq}
    I.~Antoniadis, K.~Benakli and M.~Quiros,
    \emph{Direct collider signatures of large extra dimensions},
    Phys. Lett. B \textbf{460} (1999), 176-183
    %doi:10.1016/S0370-2693(99)00764-9
    [arXiv:hep-ph/9905311 [hep-ph]].
    %206 citations counted in INSPIRE as of 06 Jul 2020
	
    %\cite{Fabbrichesi:2020wbt}
    \bibitem{Fabbrichesi:2020wbt}
    M.~Fabbrichesi, E.~Gabrielli and G.~Lanfranchi,
    \emph{The Dark Photon},
    [arXiv:2005.01515 [hep-ph]].
    
    
 	%\cite{Burgess:1998px}
	\bibitem{Burgess:1998px}
	C.~P.~Burgess, L.~E.~Ibanez and F.~Quevedo,
	\emph{Strings at the intermediate scale, or is the Fermi scale 		dual to the Planck scale?},
	Phys. Lett. B \textbf{447} (1999), 257-265
	%doi:10.1016/S0370-2693(99)00006-4
	[arXiv:hep-ph/9810535 [hep-ph]].
	%191 citations counted in INSPIRE as of 05 Jul 2020
	
	
%\cite{Anchordoqui:2020tlp}
\bibitem{Anchordoqui:2020tlp}
L.~A.~Anchordoqui, I.~Antoniadis, K.~Benakli and D.~Lust,
%``Anomalous $U(1)$ Gauge Bosons as Light Dark Matter in String Theory,''
[arXiv:2007.11697 [hep-th]].


%%%%%%%%%%%%%%%%%%% All XENON1T citation
%	
%	%\cite{Miranda:2020kwy}
%    \bibitem{Miranda:2020kwy}
%    O.~G.~Miranda, D.~K.~Papoulias, M.~Tórtola and J.~W.~F.~Valle,
%    \emph{XENON1T signal from transition neutrino magnetic moments},
%    [arXiv:2007.01765 [hep-ph]].
%    %0 citations counted in INSPIRE as of 06 Jul 2020
%
%    %\cite{Chigusa:2020bgq}
%    \bibitem{Chigusa:2020bgq}
%    S.~Chigusa, M.~Endo and K.~Kohri,
%    \emph{Constraints on electron-scattering interpretation of XENON1T excess},
%    [arXiv:2007.01663 [hep-ph]].
%    %0 citations counted in INSPIRE as of 06 Jul 2020
%
%    %\cite{Li:2020naa}
%    \bibitem{Li:2020naa}
%    T.~Li,
%    \emph{The KSVZ Axion and Pseudo-Nambu-Goldstone Boson Models for the XENON1T Excess},
%    [arXiv:2007.00874 [hep-ph]].
%    %1 citations counted in INSPIRE as of 06 Jul 2020
%
%    %\cite{Croon:2020ehi}
%    \bibitem{Croon:2020ehi}
%    D.~Croon, S.~D.~McDermott and J.~Sakstein,
%    \emph{Missing in Axion: where are XENON1T's big black holes?},
%    [arXiv:2007.00650 [hep-ph]].
%    %1 citations counted in INSPIRE as of 06 Jul 2020
%
%    %\cite{Szydagis:2020isq}
%    \bibitem{Szydagis:2020isq}
%    M.~Szydagis, C.~Levy, G.~M.~Blockinger, A.~Kamaha, N.~Parveen and G.~R.~C.~Rischbieter,
%    \emph{Investigating the XENON1T Low-Energy Electronic Recoil Excess Using NEST},
%    [arXiv:2007.00528 [hep-ex]].
%    %1 citations counted in INSPIRE as of 06 Jul 2020
%
%    %\cite{Sun:2020iim}
%    \bibitem{Sun:2020iim}
%    J.~Sun and X.~G.~He,
%    \emph{Axion-Photon Coupling Revisited},
%    [arXiv:2006.16931 [hep-ph]].
%    %3 citations counted in INSPIRE as of 06 Jul 2020
%
%    %\cite{Hryczuk:2020jhi}
%    \bibitem{Hryczuk:2020jhi}
%    A.~Hryczuk and K.~Jodłowski,
%    \emph{Self-interacting dark matter from late decays and the $H_0$ tension},
%    [arXiv:2006.16139 [hep-ph]].
%    %2 citations counted in INSPIRE as of 02 Jul 2020
%
%    %\cite{Alhazmi:2020fju}
%    \bibitem{Alhazmi:2020fju}
%    H.~Alhazmi, D.~Kim, K.~Kong, G.~Mohlabeng, J.~C.~Park and S.~Shin,
%    \emph{Implications of the XENON1T Excess on the Dark Matter Interpretation},
%    [arXiv:2006.16252 [hep-ph]].
%    %2 citations counted in INSPIRE as of 06 Jul 2020
%
%    %\cite{Cacciapaglia:2020kbf}
%    \bibitem{Cacciapaglia:2020kbf}
%    G.~Cacciapaglia, C.~Cai, M.~T.~Frandsen, M.~Rosenlyst and H.~H.~Zhang,
%    \emph{XENON1T solar axion and the Higgs boson emerging from the dark},
%    [arXiv:2006.16267 [hep-ph]].
%    %3 citations counted in INSPIRE as of 06 Jul 2020
%
%    %\cite{Gao:2020wfr}
%    \bibitem{Gao:2020wfr}
%    Y.~Gao and T.~Li,
%    \emph{Lepton Number Violating Electron Recoils at XENON1T by the $U(1)_{B-L}$ Model with Non-Standard Interactions},
%    [arXiv:2006.16192 [hep-ph]].
%    %4 citations counted in INSPIRE as of 06 Jul 2020
%
%    %\cite{Chao:2020yro}
%    \bibitem{Chao:2020yro}
%    W.~Chao, Y.~Gao and M.~j.~Jin,
%    \emph{Pseudo-Dirac Dark Matter in XENON1T},
%    [arXiv:2006.16145 [hep-ph]].
%    %4 citations counted in INSPIRE as of 06 Jul 2020
%
%    %\cite{Bhattacherjee:2020qmv}
%    \bibitem{Bhattacherjee:2020qmv}
%    B.~Bhattacherjee and R.~Sengupta,
%    \emph{XENON1T Excess: Some Possible Backgrounds},
%    [arXiv:2006.16172 [hep-ph]].
%    %5 citations counted in INSPIRE as of 06 Jul 2020
%
%    %\cite{Ge:2020jfn}
%    \bibitem{Ge:2020jfn}
%    S.~F.~Ge, P.~Pasquini and J.~Sheng,
%    \emph{Solar Neutrino Scattering with Electron into Massive Sterile Neutrino},
%    [arXiv:2006.16069 [hep-ph]].
%    %3 citations counted in INSPIRE as of 06 Jul 2020
%
%    %\cite{Dessert:2020vxy}
%    \bibitem{Dessert:2020vxy}
%    C.~Dessert, J.~W.~Foster, Y.~Kahn and B.~R.~Safdi,
%    \emph{Systematics in the XENON1T data: the 15-keV anti-axion},
%    [arXiv:2006.16220 [hep-ph]].
%    %4 citations counted in INSPIRE as of 06 Jul 2020
%
%    %\cite{DelleRose:2020pbh}
%    \bibitem{DelleRose:2020pbh}
%    L.~Delle Rose, G.~Hütsi, C.~Marzo and L.~Marzola,
%    \emph{Impact of loop-induced processes on the boosted dark matter interpretation     of the XENON1T excess},
%    [arXiv:2006.16078 [hep-ph]].
%    %4 citations counted in INSPIRE as of 06 Jul 2020
%
%    %\cite{An:2020tcg}
%    \bibitem{An:2020tcg}
%    H.~An and D.~Yang,
%    \emph{Direct detection of freeze-in inelastic dark matter},
%    [arXiv:2006.15672 [hep-ph]].
%    %4 citations counted in INSPIRE as of 06 Jul 2020
%
%    %\cite{Coloma:2020voz}
%    \bibitem{Coloma:2020voz}
%    P.~Coloma, P.~Huber and J.~M.~Link,
%    \emph{Telling Solar Neutrinos from Solar Axions When You Can't Shut Off the Sun},
%    [arXiv:2006.15767 [hep-ph]].
%    %4 citations counted in INSPIRE as of 06 Jul 2020
%
%    %\cite{McKeen:2020vpf}
%    \bibitem{McKeen:2020vpf}
%    D.~McKeen, M.~Pospelov and N.~Raj,
%    \emph{Hydrogen portal to exotic radioactivity},
%    [arXiv:2006.15140 [hep-ph]].
%    %6 citations counted in INSPIRE as of 06 Jul 2020
%
%    %\cite{Zioutas:2020cul}
%    \bibitem{Zioutas:2020cul}
%    K.~Zioutas, G.~Cantatore, M.~Karuza, A.~Kryemadhi, M.~Maroudas and Y.~K.~Semertzidis,
%    \emph{Response-suggestion to The XENON1T excess: an overlooked dark matter signature?},
%    [arXiv:2006.16907 [hep-ph]].
%    %2 citations counted in INSPIRE as of 06 Jul 2020
%
%    %\cite{Dent:2020jhf}
%    \bibitem{Dent:2020jhf}
%    J.~B.~Dent, B.~Dutta, J.~L.~Newstead and A.~Thompson,
%    \emph{Inverse Primakoff Scattering as a Probe of Solar Axions at Liquid Xenon Direct Detection Experiments},
%    [arXiv:2006.15118 [hep-ph]].
%    %10 citations counted in INSPIRE as of 06 Jul 2020
%
%    %\cite{DeRocco:2020xdt}
%    \bibitem{DeRocco:2020xdt}
%    W.~DeRocco, P.~W.~Graham and S.~Rajendran,
%    \emph{Exploring the robustness of stellar cooling constraints on light particles},
%    [arXiv:2006.15112 [hep-ph]].
%    %7 citations counted in INSPIRE as of 06 Jul 2020
%
%    %\cite{Bloch:2020uzh}
%    \bibitem{Bloch:2020uzh}
%    I.~M.~Bloch, A.~Caputo, R.~Essig, D.~Redigolo, M.~Sholapurkar and T.~Volansky,
%    \emph{Exploring New Physics with O(keV) Electron Recoils in Direct Detection Experiments},
%    [arXiv:2006.14521 [hep-ph]].
%    %14 citations counted in INSPIRE as of 06 Jul 2020
%
%    %\cite{Chala:2020pbn}
%    \bibitem{Chala:2020pbn}
%    M.~Chala and A.~Titov,
%    \emph{One-loop running of dimension-six Higgs-neutrino operators and implications of a large neutrino dipole moment},
%    [arXiv:2006.14596 [hep-ph]].
%    %7 citations counted in INSPIRE as of 06 Jul 2020
%
%    %\cite{Lindner:2020kko}
%    \bibitem{Lindner:2020kko}
%    M.~Lindner, Y.~Mambrini, T.~B.~de Melo and F.~S.~Queiroz,
%    \emph{XENON1T Anomaly: A Light $Z^\prime$},
%    [arXiv:2006.14590 [hep-ph]].
%    %14 citations counted in INSPIRE as of 06 Jul 2020
%
%    %\cite{Budnik:2020nwz}
%    \bibitem{Budnik:2020nwz}
%    R.~Budnik, H.~Kim, O.~Matsedonskyi, G.~Perez and Y.~Soreq,
%    \emph{Probing the relaxed relaxion and Higgs-portal with S1 $\&$ S2},
%    [arXiv:2006.14568 [hep-ph]].
%    %9 citations counted in INSPIRE as of 06 Jul 2020
%
%    %\cite{Gao:2020wer}
%    \bibitem{Gao:2020wer}
%    C.~Gao, J.~Liu, L.~T.~Wang, X.~P.~Wang, W.~Xue and Y.~M.~Zhong,
%    \emph{Re-examining the Solar Axion Explanation for the XENON1T Excess},
%    [arXiv:2006.14598 [hep-ph]].
%    %17 citations counted in INSPIRE as of 06 Jul 2020
%
%
%    %\cite{Baryakhtar:2020rwy}
%    \bibitem{Baryakhtar:2020rwy}
%    M.~Baryakhtar, A.~Berlin, H.~Liu and N.~Weiner,
%    \emph{Electromagnetic Signals of Inelastic Dark Matter Scattering},
%    [arXiv:2006.13918 [hep-ph]].
%    %18 citations counted in INSPIRE as of 06 Jul 2020
%
%    %\cite{Bramante:2020zos}
%    \bibitem{Bramante:2020zos}
%    J.~Bramante and N.~Song,
%    \emph{Electric But Not Eclectic: Thermal Relic Dark Matter for the XENON1T Excess},
%    [arXiv:2006.14089 [hep-ph]].
%    %12 citations counted in INSPIRE as of 06 Jul 2020
%
%    %\cite{Jho:2020sku}
%    \bibitem{Jho:2020sku}
%    Y.~Jho, J.~C.~Park, S.~C.~Park and P.~Y.~Tseng,
%    \emph{Gauged Lepton Number and Cosmic-ray Boosted Dark Matter for the XENON1T Excess},
%    [arXiv:2006.13910 [hep-ph]].
%    %15 citations counted in INSPIRE as of 06 Jul 2020
%
%    %\cite{Gelmini:2020xir}
%    \bibitem{Gelmini:2020xir}
%    G.~B.~Gelmini, V.~Takhistov and E.~Vitagliano,
%    \emph{Scalar Direct Detection: In-Medium Effects},
%    [arXiv:2006.13909 [hep-ph]].
%    %3 citations counted in INSPIRE as of 04 Jul 2020
%
%
%    %\cite{Primulando:2020rdk}
%    \bibitem{Primulando:2020rdk}
%    R.~Primulando, J.~Julio and P.~Uttayarat,
%    \emph{Collider Constraints on a Dark Matter Interpretation of the XENON1T Excess},
%    [arXiv:2006.13161 [hep-ph]].
%    %18 citations counted in INSPIRE as of 06 Jul 2020
%
%    %\cite{Khan:2020vaf}
%    \bibitem{Khan:2020vaf}
%    A.~N.~Khan,
%    \emph{Can nonstandard neutrino interactions explain the XENON1T spectral excess?},
%    [arXiv:2006.12887 [hep-ph]].
%    %20 citations counted in INSPIRE as of 06 Jul 2020
%
%    %\cite{Cao:2020bwd}
%    \bibitem{Cao:2020bwd}
%    Q.~H.~Cao, R.~Ding and Q.~F.~Xiang,
%    \emph{Exploring for sub-MeV Boosted Dark Matter from Xenon Electron Direct Detection},
%    [arXiv:2006.12767 [hep-ph]].
%    %19 citations counted in INSPIRE as of 06 Jul 2020
%
%    %\cite{Robinson:2020gfu}
%    \bibitem{Robinson:2020gfu}
%    A.~E.~Robinson,
%    \emph{XENON1T observes tritium},
%    [arXiv:2006.13278 [hep-ex]].
%    %13 citations counted in INSPIRE as of 06 Jul 2020
%
%    %\cite{Lee:2020wmh}
%    \bibitem{Lee:2020wmh}
%    H.~M.~Lee,
%    \emph{Exothermic Dark Matter for XENON1T Excess},
%    [arXiv:2006.13183 [hep-ph]].
%    %19 citations counted in INSPIRE as of 06 Jul 2020
%
%    %\cite{Paz:2020pbc}
%    \bibitem{Paz:2020pbc}
%    G.~Paz, A.~A.~Petrov, M.~Tammaro and J.~Zupan,
%    \emph{Shining dark matter in Xenon1T},
%    [arXiv:2006.12462 [hep-ph]].
%    %23 citations counted in INSPIRE as of 06 Jul 2020
%
%    %\cite{Buch:2020mrg}
%    \bibitem{Buch:2020mrg}
%    J.~Buch, M.~A.~Buen-Abad, J.~Fan and J.~S.~C.~Leung,
%    \emph{Galactic Origin of Relativistic Bosons and XENON1T Excess},
%    [arXiv:2006.12488 [hep-ph]].
%    %23 citations counted in INSPIRE as of 06 Jul 2020
%
%    %\cite{AristizabalSierra:2020edu}
%    \bibitem{AristizabalSierra:2020edu}
%    D.~Aristizabal Sierra, V.~De Romeri, L.~J.~Flores and D.~K.~Papoulias,
%    \emph{Light vector mediators facing XENON1T data},
%    [arXiv:2006.12457 [hep-ph]].
%    %27 citations counted in INSPIRE as of 06 Jul 2020
%
%    %\cite{Bell:2020bes}
%    \bibitem{Bell:2020bes}
%    N.~F.~Bell, J.~B.~Dent, B.~Dutta, S.~Ghosh, J.~Kumar and J.~L.~Newstead,
%    \emph{Explaining the XENON1T excess with Luminous Dark Matter},
%    [arXiv:2006.12461 [hep-ph]].
%    %25 citations counted in INSPIRE as of 06 Jul 2020
%
%    %\cite{Dey:2020sai}
%    \bibitem{Dey:2020sai}
%    U.~K.~Dey, T.~N.~Maity and T.~S.~Ray,
%    \emph{Prospects of Migdal Effect in the Explanation of XENON1T Electron Recoil Excess},
%    [arXiv:2006.12529 [hep-ph]].
%    %20 citations counted in INSPIRE as of 06 Jul 2020
%
%    %\cite{Chen:2020gcl}
%    \bibitem{Chen:2020gcl}
%    Y.~Chen, J.~Shu, X.~Xue, G.~Yuan and Q.~Yuan,
%    \emph{Sun Heated MeV-scale Dark Matter and the XENON1T Electron Recoil Excess},
%    [arXiv:2006.12447 [hep-ph]].
%    %26 citations counted in INSPIRE as of 06 Jul 2020
%
%    %\cite{DiLuzio:2020jjp}
%    \bibitem{DiLuzio:2020jjp}
%    L.~Di Luzio, M.~Fedele, M.~Giannotti, F.~Mescia and E.~Nardi,
%    \emph{Solar axions cannot explain the XENON1T excess},
%    [arXiv:2006.12487 [hep-ph]].
%    %31 citations counted in INSPIRE as of 06 Jul 2020
%
%    %\cite{Du:2020ybt}
%    \bibitem{Du:2020ybt}
%    M.~Du, J.~Liang, Z.~Liu, V.~Tran and Y.~Xue,
%    \emph{On-shell mediator dark matter models and the Xenon1T anomaly},
%    [arXiv:2006.11949 [hep-ph]].
%    %24 citations counted in INSPIRE as of 06 Jul 2020
%
%    %\cite{Su:2020zny}
%    \bibitem{Su:2020zny}
%    L.~Su, W.~Wang, L.~Wu, J.~M.~Yang and B.~Zhu,
%    \emph{Atmospheric Dark Matter from Inelastic Cosmic Ray Collision in Xenon1T},
%    [arXiv:2006.11837 [hep-ph]].
%    %22 citations counted in INSPIRE as of 06 Jul 2020
%
%    %\cite{Bally:2020yid}
%    \bibitem{Bally:2020yid}
%    A.~Bally, S.~Jana and A.~Trautner,
%    \emph{Neutrino self-interactions and XENON1T electron recoil excess},
%    [arXiv:2006.11919 [hep-ph]].
%    %25 citations counted in INSPIRE as of 06 Jul 2020
%
%    %\cite{Harigaya:2020ckz}
%    \bibitem{Harigaya:2020ckz}
%    K.~Harigaya, Y.~Nakai and M.~Suzuki,
%    \emph{Inelastic Dark Matter Electron Scattering and the XENON1T Excess},
%    [arXiv:2006.11938 [hep-ph]].
%    %24 citations counted in INSPIRE as of 06 Jul 2020
%
%    %\cite{Boehm:2020ltd}
%    \bibitem{Boehm:2020ltd}
%    C.~Boehm, D.~G.~Cerdeno, M.~Fairbairn, P.~A.~N.~Machado and A.~C.~Vincent,
%    \emph{Light new physics in XENON1T},
%    [arXiv:2006.11250 [hep-ph]].
%    %38 citations counted in INSPIRE as of 06 Jul 2020
%
%    %\cite{Fornal:2020npv}
%    \bibitem{Fornal:2020npv}
%    B.~Fornal, P.~Sandick, J.~Shu, M.~Su and Y.~Zhao,
%    \emph{Boosted Dark Matter Interpretation of the XENON1T Excess},
%    [arXiv:2006.11264 [hep-ph]].
%    %35 citations counted in INSPIRE as of 06 Jul 2020
%
%    %\cite{Amaral:2020tga}
%    \bibitem{Amaral:2020tga}
%    D.~W.~P.~Amaral, do., D.~G.~Cerdeno, P.~Foldenauer and E.~Reid,
%    \emph{Solar neutrino probes of the muon anomalous magnetic moment in the gauged $U(1)_{L_\mu-L_\tau}$},
%    [arXiv:2006.11225 [hep-ph]].
%    %12 citations counted in INSPIRE as of 06 Jul 2020
%
%
%    %\cite{Kannike:2020agf}
%    \bibitem{Kannike:2020agf}
%    K.~Kannike, M.~Raidal, H.~Veermäe, A.~Strumia and D.~Teresi,
%    \emph{Dark Matter and the XENON1T electron recoil excess},
%    [arXiv:2006.10735 [hep-ph]].
%    %41 citations counted in INSPIRE as of 06 Jul 2020
%
%    %\cite{OHare:2020wum}
%    \bibitem{OHare:2020wum}
%    C.~A.~J.~O'Hare, A.~Caputo, A.~J.~Millar and E.~Vitagliano,
%    \emph{Axion helioscopes as solar magnetometers},
%    [arXiv:2006.10415 [astro-ph.CO]].
%    %5 citations counted in INSPIRE as of 02 Jul 2020
%
%    %\cite{Takahashi:2020bpq}
%    \bibitem{Takahashi:2020bpq}
%    F.~Takahashi, M.~Yamada and W.~Yin,
%    \emph{XENON1T anomaly from anomaly-free ALP dark matter and its implications for stellar cooling anomaly},
%    [arXiv:2006.10035 [hep-ph]].
%    %40 citations counted in INSPIRE as of 06 Jul 2020
%
%    %\cite{Smirnov:2020zwf}
%    \bibitem{Smirnov:2020zwf}
%    J.~Smirnov and J.~F.~Beacom,
%    \emph{Co-SIMP Miracle},
%    [arXiv:2002.04038 [hep-ph]].
%    %7 citations counted in INSPIRE as of 02 Jul 2020

\end{thebibliography}
\end{document}